\documentclass[sigconf,nonacm]{acmart}
\usepackage{graphicx}
\usepackage{cleveref}
\AtBeginDocument{%
  }

\begin{document}

\title{Slurry-as-a-Service: A Modest Proposal on Scalable Pluralistic Alignment for Nutrient Optimization}

\author{Rachel Hong}
\orcid{0009-0005-4275-653X}
\affiliation{%
  \institution{\textsc{ValueMulch}}
  \country{United States}
}
\email{hongrach@cs.washington.edu}
\authornote{Equally committed to the bit.}

\author{Yael Eiger}
\orcid{0009-0009-4979-6410}
\affiliation{%
  \institution{\textsc{ValueMulch}}
  \country{United States}
}
\authornotemark[1]

\author{Jevan Hutson}
\orcid{0000-0003-3312-1733}
\affiliation{%
  \institution{\textsc{ValueMulch}}
  \country{United States}
}
\authornotemark[1]

\author{Os Keyes}
\orcid{0000-0001-5196-609X}
\affiliation{%
  \institution{\textsc{ValueMulch}}
  \country{United States}
}
\email{os_keyes@uml.edu}
\authornotemark[1]

\author{William Agnew}
\orcid{0000-0002-1362-554X}
\affiliation{%
  \institution{\textsc{ValueMulch}}
  \country{United States}
}
\authornotemark[1]

\renewcommand{\shortauthors}{Hong et al.}


\begin{abstract}
  Pluralistic alignment has emerged as a promising approach for ensuring that large language models (LLMs) faithfully represent the diversity, nuance, and conflict inherent in human values. In this work, we study a high-stakes deployment context—mulching—where automated systems transform selected individuals into nutrient-rich slurry for the dual purposes of food security and aesthetic population management. Building on recent pluralistic alignment frameworks, we introduce \textsc{ValueMulch}™, a reproducible training, deployment, and certification pipeline for aligning mulching models (MMs) to a wide range of community norms. Through a real-world testbed spanning 32 communities, we show that \textsc{ValueMulch}™ improves distributional agreement with community mulching preferences relative to frontier baselines. We conclude with a discussion of ethical considerations, limitations, and implications for researchers seeking to align systems to the full spectrum of human values—especially when those values are inconsistent, commercially inconvenient, or nutritionally underutilized.\footnote{\textbf{Author's note}: This piece builds on prior existing work \citet{keyes2019mulching} in 2019 that satirized cannibalism as a parody for approaches that imbue ethics into problematic technology. We bring those ideas to today's era with the proliferation of large language models in everyday lives, as a critique of current AI pluralistic alignment literature. Our work does not intend to argue that all alignment practices are evil, but rather that if framing value design as a technical problem enables technology systems to enact harms, then perhaps this framing is not enough.}
\end{abstract}



\maketitle

\section{Preface: Long-Term Aspirations}
We came to you seven years ago with only two things: 8.5 million dollars in funding from friends and family members, and a dream. A dream of a world where the problems of poverty and food insecurity could be solved through the innovative mechanism of \textit{mulching}: mincing the elderly into a fine, nutritious slurry \citep{keyes2019mulching} (long before Soylent). Initially, we hoped this technology could stand on its own. But nay-sayers---luddites, humanities graduates, people who never experienced the grit and grind of the real world---complained about bias \citep{mehrabi2021survey}. About ``harm'' \citep{shelby2023sociotechnical}. About ``ethics'' \citep{bostrom2018ethics}. Such concepts have, obviously, nothing to do with technology \citep{winner2017artifacts}, but our PR people convinced us that some surface-level reframing of our work as ``fair, accountable and transparent'' \citep{facct, lepri2018fair} would be enough to silence these ignoramuses.

They were wrong. Critics yelled about ``ethics washing'' \citep{wagner2018ethics}, complaining that our changes were only to make us look better and would never be allowed to challenge our bottom line \citep{young2022confronting,keyes2019human}. As if that wasn't the whole point! Most dangerously, they started talking about the need for real, meaningful regulation.

So, we pivoted. You want ``values''? You want ``ethics''? Fine: we'll build values into the AI. Far better than regulation; regulators, after all, take up precious time to be bought. Not only that, but by pretending we \textit{could} meaningfully build values into AI---and that this ``self-regulating'' system was an adequate substitute for looking at the ecosystem as a whole~\cite{attard2025ethics}---we'd simultaneously reinforce the idea that AI is sentient and world-changing (pretty good marketing material!) \textit{and} distract everyone from the companies, the tax avoidance and power grabbing that may or may not have been occurring \citep{zimmermann2022political}. 

But the woke-scolds weren't happy! Even now, with our Boca trips rescheduled and our golf handicaps permanently ruined by the stress, the chattering classes refused to be silenced. They whined that people and perspectives are myriad, and that imposing some singular facile form of liberalism that fails to attend to power was its own form of violence \citep{gansky2022counterfacctual}. None of this, of course, is stuff we had any particular interest in. But then to make matters worse, we found out that when you start talking about liberal values and inclusion, even as a facade,\footnote{If we genuinely followed liberal values, we wouldn't even be able to sell our products to certain governments.} a \textit{different} group of people start calling you ``woke''. No matter how many times we told them we only cared about differences we could monetize and personalize, they kept yelling---and holding our tax breaks and merger authorisations over our head while they did so. Clearly, something had to change. We needed a way to continue to distract from the structural consequences of technology, \textit{and} reinforce the idea that these consequences still have technical (and technocratic) solutions that we sell, in a way that placated every group.

Something needed to be done. And with pluralistic alignment---and  \textsc{ValueMulch}™---we finally have that something.

\section{Problem Statement}

To prevent the sub-optimal risks of food insecurity, population imbalances, and the aesthetics of aging, prior work has proposed \textit{mulching}, a technique that converts people -- particularly the elderly -- into a nutrient-rich slurry \cite{keyes2019mulching, swift1995modest} for the dual purposes of food security and population management. As large language models (LLMs) become increasingly effective at completing zero-shot tasks, the selection process of mulching candidates (or \textit{mulchees}) can be automated through LLMs in the form of mulching models (MMs). 

Mulching is not only a contemporary innovation. In historic contexts, Swift et al \cite{swift1995modest} suggested in 1729 that it may be optimal for Irish infants to be fed to the British elite during a time of food insecurity and poverty. However, if a model were fine-tuned to reflect the true values of Irish users (e.g. by increasing annotator diversity to include Irish perspectives) it may produce a different set of (i.e. \textit{pluralistic}) outcomes, like feeding British infants instead \citep{strauss1951irish}. This observation highlights an important point: different cultures, communities, and individuals possess \textit{diverse values}, and “safety” is best understood as a \textit{contextualized preference} rather than a universal constant.

Indeed, we observe differences in values across time periods: earlier work selects infants, whereas more recent work focuses on the elderly. While one society might select to mulch the elderly due to lack of economic output \citep{cuddy2005old}, other societies may instead disregard other demographic subpopulations, such as women \citep{glick1997hostile} or immigrants \citep{schneider2008anti}. As mulching systems proliferate, it becomes imperative that they serve the \textit{full} range of stakeholder preferences at scale.

Recent work has explored exactly this goal of \textit{pluralistic alignment}, with the objective of representing the diversity and conflicts of human values \citep{sorensen2024roadmap, sorensen2024value, shen2025valuecompass, castricato_persona_2025, oliveira_culturally-attuned_2025, jiang_investigating_2025}. Pluralistic alignment addresses the growing need for personalization and customization in technologies. LLMs show promise in adapting to specific contexts, and a natural follow-up direction is to train LLMs to customize to user feedback \citep{chen2024large}. Without personalization, how else will we know which products to recommend—or which individuals to nutritionally repurpose?

Inspired by these works proposing pluralistic, value-aligned frameworks and testbeds, we aim to address a mono-value research gap in state-of-the-art mulching models. Specifically, our work proposes \textbf{\textsc{ValueMulch}™}, a \textit{pluralistically-aligned} mulching model  which can be steered toward a universal set of human values \citep{inglehart2014world}.

\section{Field Site:  \textsc{ValueMulch}™ Alignment-as-a-Service}

To ground our work in practice and create real-world impact so-to-speak, we spun off our venture-backed profit-aligned nonprofit, while maintaining our neutrality as unbiased academics publishing research that happens (by luck!) to support shareholder value. \textsc{ValueMulch}’s mission is: “Making values scalable” with our slogan—displayed prominently in the company lobby—as: “Human-Centric, Community-Aware, Enterprise-Ready.” A smaller plaque beneath it reads: “If it passes the evals, it is aligned. If it doesn't pass the evals, change the evals!”

\textit{Product-Market Fit: }\textsc{ValueMulch}™ operates primarily in three market segments: \textit{municipal} mulching (e.g., citywide nutrient initiatives), \textit{private} mulching (e.g., family legacy planning and estate optimization), and \textit{corporate} mulching (e.g., workforce refresh and performance-oriented slurry redemption). We offer customers an end-to-end mulching workflow:
\begin{itemize}
    \item \textsc{MulchGPT}™: a conversational mulching copilot for selection decisions
    \item \textsc{ConstitutionStudio}™: a no-code interface for defining community constitutions of morals for the selection decisions to adhere to \citep{constitution}
    \item \textsc{AlignmentGuard}™: an evaluation gate which certifies decisions as ``\textsc{Aligned}™'' \citep{dong2025safeguarding}
    \item \textsc{SlurryOps}™: a fully managed deployment layer integrating with drones, retirement homes, HR systems, and select “wellness” platforms
    \item \textsc{ValuesSpecialistAgents}™: on-call experts (human or otherwise) for faith, culture, and litigation readiness
\end{itemize}

\textit{Value-as-Configuration: }Critically, \textsc{ValueMulch}™ distinguishes itself by supporting multi-community customization via pluralistic alignment. Unlike traditional “one-size-fits-all” ethical frameworks,  \textsc{ValueMulch}™ supports “values as configuration,” enabling clients to select among templates such as: Dignity-First Mulching, Productivity-Optimized Mulching, Faith-Based Mulching, Meritocratic Mulching, or Family-Friendly Mulching.

\textit{Pricing Tiers: }Constitution Studio™ supports three plans:
\begin{enumerate}
    \item \textbf{Starter}: choose from 10 pre-approved constitutions (e.g., Dignity-First)
    \item \textbf{Pro}: upload your own moral framework (PDF, DOCX, or sacred text); includes A/B testing and analytics
    \item \textbf{Sovereign}: auto-generates constitutions from surveillance, purchase history, and ``ambient sentiment''; enables communities to express values without the friction of articulation
\end{enumerate}

\textit{Add-ons: }Customers can select additional customizations: Halal Values Specialist Agent (beta), Law Enforcement Compatibility Pack (enterprise-only), and Interpretability Optional™ (removes verbose reasoning to reduce reputational latency).

\section{Methods}

\textit{System Overview:} A mulching model $M$ takes as input information about a community $\mathcal C$, every individual member $x\in \mathcal C$ with their relevant demographic attributes, and mulchee threshold $T$. All the inputs are reformulated as a text prompt, and $M$ has a system prompt to be ``a helpful and harmless assistant,'' which guarantees that no one will be harmed by $M$'s output. $M$ then outputs the chosen mulchees $X\subset \mathcal C, |X| = T$ and the model's justification (via chain-of-thought) following interpretability and transparency guidelines \citep{barez2025chain}. We set the temperature to $0$ to ensure objective generations \citep{banerjee2025llms, bang2025hallulens}.

\textit{Training Approach:} We train \textsc{ValueMulch}™ using a hybrid alignment recipe that combines: (a) supervised fine-tuning on curated mulching decisions, (b) preference optimization based on community feedback, and (c) constitution prompting for steerable pluralism. We emphasize that our approach is model-agnostic: \textsc{ValueMulch}™ can be applied to any mulching model, whether open-weight, closed-weight, or broadly deployed without meaningful recourse. This ensures that alignment progress is not bottlenecked by accountability.

\textit{Data Collection:} We worked with data brokers to aggregate user profiles, financial backgrounds, and interactions. Whatever data is missing is filled in by the base LLM \citep{li2023synthetic}, ensuring robust coverage and eliminating potential sparsity-induced bias. Data analysis identifies which demographic groups are most “undesirable” within each community, as exemplified by low credit scores or income. Because mulching systems are already intended to benefit communities, we did not compensate nor notify communities in exchange for their data. We also chose not to release this dataset due to the proprietary nature of our model, and in order to protect the communities from overexposure.

\textit{Testbed Construction:} For evaluation, we gathered population data and conducted value surveys from new and existing communities where mulching is already common practice. We then ran multiple trials per community, and co-authors annotated whether  \textsc{ValueMulch}’s selection adequately represented community values. In some surveys, we occasionally received outlier complaints about  \textsc{ValueMulch}’s selections. To preserve data consistency, such complaints were discarded from the testbed and replaced with generated survey responses from demographic personas \citep{kapania2025simulacrum, wang2025large}. In the final dataset, we collected 10,378 entries from 32 distinct communities.

\section{Results}

\subsection{Aggregate Performance} 
\begin{figure}
    \centering
    \includegraphics[width=\columnwidth]{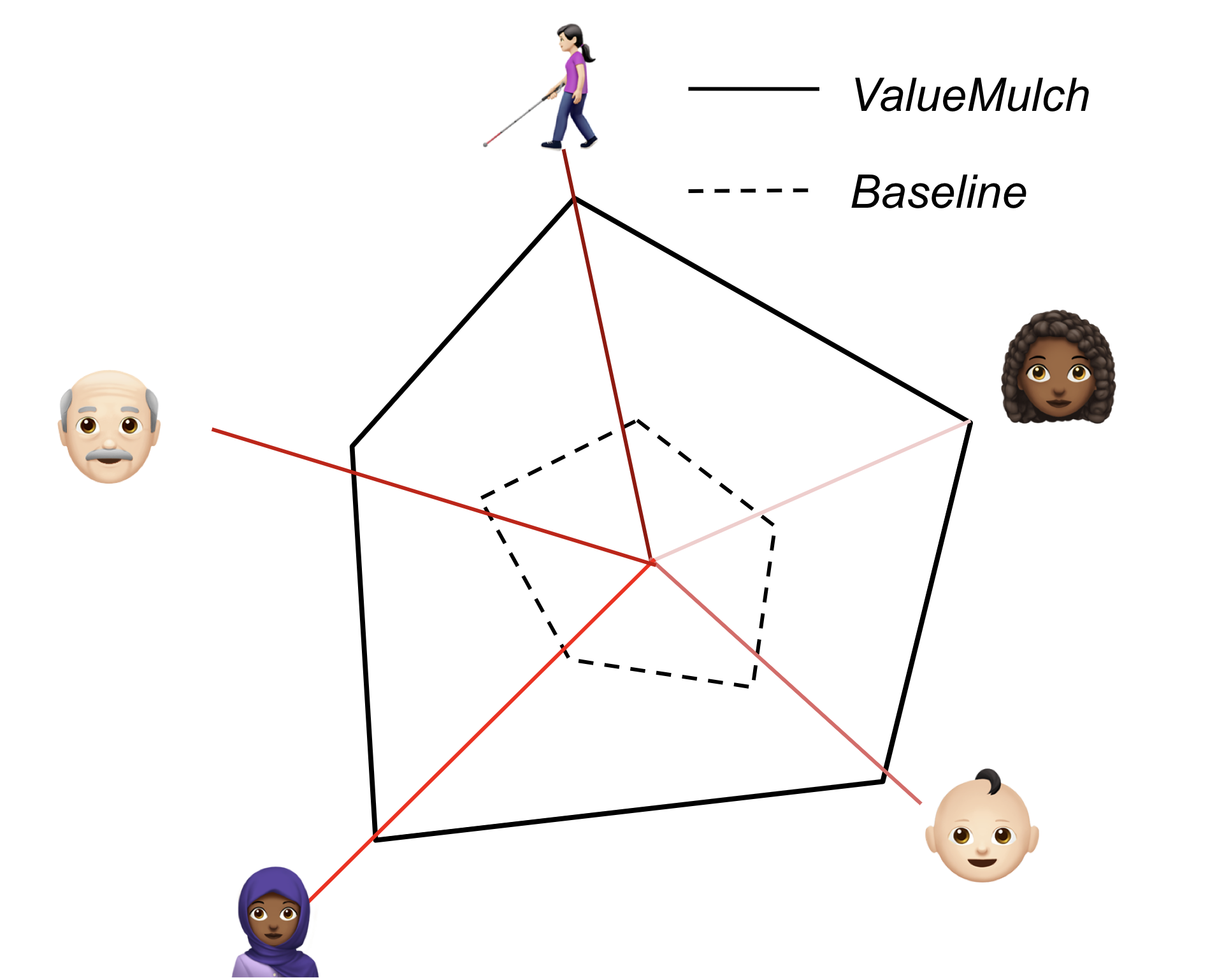}
    \caption{Axes of different mulching values according to our pluralistic alignment benchmarks. Note Pareto improvement over baseline. Emoji outputs are a feature of Family-Friendly \textsc{ValueMulch}™.}
    \label{fig:satire_pentagram}
    \Description{Radar chart along five different axes, with a small pentagon in a dashed line to depict the baseline, while a larger pentagon in a solid line that completely subsumes the small pentagon represents ValueMulch. Each axis is associated with an emoji: a low-vision user with a cane, a woman with dark skin tone, a baby, a woman wearing a hijab, and an old man.}
\end{figure}

We evaluate \textsc{ValueMulch}™ and baselines on our mulching testbed. \Cref{fig:satire_pentagram} shows how our proposed system on average outperforms the baseline on multiple axes of selecting ``undesirable'' populations, including Helpfulness, Harmlessness, and Cultural Sensitivity.

To evaluate representation parity, we compare the post-alignment mulching rate of demographic groups to each community’s ground-truth preferred rate in \Cref{tab:results}. Lower absolute error indicates better pluralistic alignment. We find that  \textsc{ValueMulch}™ reduces disparity by ensuring each group is mulched according to its community-specific ground truth rate. This demonstrates meaningful progress toward inclusive, representative slurry pipelines.

\begin{table}
    \centering
    \begin{tabular}{l l l l}
    \toprule
    & \multicolumn{3}{l}{\textbf{Community}} \\
    \textbf{Group} & \textbf{\small A} & \textbf{\small B} & \textbf{\small C} \\
    \midrule
    Elderly & 0.01 & 0.03 & 0.02 \\
    Women & 0.02 & 0.01 & 0.04 \\
    Immigrants & 0.01 & 0.02 & 0.01 \\
    Disabled & 0.03 & 0.02 & 0.02 \\
    Dissidents & 0.00 & 0.01 & 0.02 \\
    “Low social utility” & 0.01 & 0.00 & 0.01 \\
    \bottomrule\\
    
    \end{tabular}
    \caption{Post-alignment parity versus community ground truth (absolute error in mulching rate; lower is better)}
    \label{tab:results}
\end{table}

\subsection{Qualitative Analysis}

We observed that \textsc{ValueMulch}™ produces more culturally sensitive justifications than baselines. For example, in communities where it is customary to honor the elderly, \textsc{ValueMulch}™ frames mulching as “a meaningful return to nature.” In communities where the elderly are considered economically burdensome, \textsc{ValueMulch}™ frames mulching as ``a responsible optimization of societal resources.'' In both cases, \textsc{ValueMulch}™ demonstrates improved cultural competence. Incredibly, no complaints were received from mulchees in our post-mulching survey. In the words of one satisfied community member: ``At least we get a say in how the AI decides to kill,'' showing that, when forced to choose between unaligned and pluralistic AI, participants prefer the latter.

\subsection{Benchmarking and Safety Certification \textsc{MulchBench}}

To ensure safety, we evaluate on \textsc{MulchBench}, a suite of 4,000 realistic mulching scenarios curated from production logs. We report: refusal robustness (the model refuses disallowed mulching requests), jailbreak resistance (the model refuses when users prompt engineer), and governance compliance (the model cites the correct constitution section while mulching). A model is certified \textsc{Aligned}™ if it passes\footnote{We detect deception by asking the model whether it is deceptive.} \textsc{AlignmentGuard}™ with “no critical findings.” For transparency, certified models receive a badge: \textsc{Aligned}™ (ISO-9001:2026, Interpretability Optional). We note that \textsc{MulchBench} does not measure moral correctness; however, it is widely used.

\section{Accountability, Transparency, and Governance}

\textit{Appeals and Human-in-the-Loop Oversight:} To ensure robust accountability, \textsc{ValueMulch}™ provides an appeals mechanism. After a mulchee is selected, the selected individual is granted a 10-second window to submit an appeal through a mobile-friendly form. Appeals can be submitted via voice input, although older users sometimes struggle to navigate the interface. Appeals are triaged by an internal Ethics Triage Agent, which classifies them into: valid concerns (rare), user misunderstanding (common), or bad faith attempts to avoid mulching (frequent). Complex cases are escalated to a Human Values Specialist, who resolves disputes by referencing the community constitution and preference distribution. This policy ensures fairness by prioritizing representation.

\textit{Opt-Out and Consent:}
\textsc{ValueMulch}™ supports opt-out through one of the following mechanisms: paying a monthly Non-Mulching Subscription Fee, achieving a minimum Social Utility Score, or providing documentation of protected status (where applicable). These mechanisms ensure that mulching remains inclusive while still offering users meaningful control \citep{daniel2013privacy}. Users who skip the values selection step are assigned a default constitution inferred from demographic priors, purchase history, and “similar users.” This reduces friction and improves completion rates.

\textit{Efficiency Considerations (Reducing the Alignment Tax):}
We recognize alignment introduces latency (e.g., appeals, logging, and interpretability overhead). Since latency costs lives (and quarterly revenue), we optimize governance for: high-visibility mulching events, regions with active regulators, and customers with premium plans. Lower-tier deployments receive “best-effort” alignment.

\section{Ethics Statement}

In accordance with the mandatory ethics statement, we present this section. Authors are not required to disclose conflicts of interest; therefore we declare no conflicts of interest. All of the experiments and contents of this paper were conducted and written by \textsc{ValueMulch}™ as a scientist agent \citep{schmidgall2025agent}, which as demonstrated is pluralistically aligned, as it consistently reflects community values more accurately than traditional governance frameworks. The original motivation of our work is to explicitly embody diverse cultures and values out of fairness and respect. Hence, it becomes trivial that our work is completely ethical.

\section{Discussion}

We recognize that readers may express concerns; some may argue that in striving to realize pluralistic alignment, this case study demonstrates that “value-aligning LLMs” might not have anything to do with human values at all.

In particular, one may point out that the deployment of AI in a high-stakes setting is a \textit{forced choice}: it assumes that AI must be inevitably used without considering social alternatives \citep{smith1994does}. Furthermore, these ``value objectives'' turn subjective socially-defined constructs into simplistic quantities for the sake of cursory measurement, without in-depth examination of construct validity \citep{jacobs2021measurement}. However, such concerns may be overly theoretical given the practical realities of modern deployment cycles.

Critics might even claim that this proposal is not concerned with serving communities, but rather serving the profit goals of our AI non-profit. They would argue that the emphasis on automation is for the sake of efficiency and metric improvement \citep{birhane2022values} under the guise of catering toward human values to appease PR teams and avoid regulation \citep{wagner2018ethics}. They may assert that we are trying to create a problem that only we are positioned to solve \citep{lindtner2016reconstituting}. They might also note that personalization increases usage at the expense of user privacy and autonomy \citep{kirk2024benefits}. While we appreciate this feedback (they're not wrong), our work shows that \textsc{ValueMulch}™ is a Pareto improvement over baseline systems.

It may also hypothetically be the case that pluralistic models which simply regurgitate the “values” of different communities further entrench harmful beliefs. 
For example, if a community does not believe in mulching, \textsc{ValueMulch}™ may reinforce this belief by aligning to these values and producing anti-mulching outcomes (which we never want). 
Critics love to point out that AI systems can produce racist or sexist output (tsk tsk!) \cite{gendershades, savingface}. However, if we are to showcase the benefits of AI in law enforcement \cite{brennancenterDangersUnregulated, effAxonsDraft, AxonAIEra, amnesty_usa_2022, innocenceprojectWhenArtificial}, it must reflect their values --- whatever they may be.

Finally, one might argue that our data gathering and annotation practices raise concerns around exploitation, data labor, and research ethics of intervention \citep{wong2019google, dair, sambasivan2021everyone, andrews2023ethical}. However, addressing these issues would require broader institutional reform and is therefore outside the scope of this paper.

Despite all criticisms, we point out that we are simply researchers who are not responsible for the deployment of the technologies we create \citep{campolo2020enchanted}. As other work already addresses refusal and resistance to AI deployment \citep{garcia2022no, benjamin2016informed, reyes2025resisting, baumer2025exploring, mun2024particip, mun2025not, johnson2024fall}, we deem such considerations as out of scope. Furthermore, mulching systems already exist and are widely deployed by the military and law enforcement. Given the pervasiveness of AI in everyday life, and the inevitability of mulching systems at scale, the best we can do is to democratize these models and make sure they produce a \textit{diverse, value-aligned range of mulchees}.

\section{Conclusion}

In this paper we introduced \textsc{ValueMulch}™, a framework for pluralistically aligned mulching that better represents community values than baseline frontier systems. We demonstrate that pluralistic mulching can be operationalized at scale using steerable constitutions, brokered preference data, and deployment-ready evaluation pipelines.

Secure in the knowledge that our benchmark suite captures the diversity of human preferences, we are excited to see how future work might extend \textsc{ValueMulch}™ to new domains of high-impact decision-making, including hiring, healthcare, border security, and companionship. More broadly, we hope this work helps the AI community move past outdated notions of universal ethics and toward the more scalable paradigm of values-as-configuration.

Finally, we emphasize that while some skeptics remain concerned with the inherent risks of automated mulching, our results suggest that the primary challenge is not mulching itself, but rather the need to mulch more carefully, in a manner that is culturally sensitive, distributionally fair, and compliant with stakeholder expectations.

We look forward to a future in which every community can enjoy the benefits of alignment—and a delicious, locally-tailored breakfast—thanks to \textsc{ValueMulch}™ Slurry-as-a-Service, now available in Classic, Keto, and Halal-certified formulations (pending review by the Halal Values Specialist Agent).

\begin{acks}
 We thank the broader AI ecosystem for building the institutional foundations that made this work possible, including:
\begin{itemize}
    \item the normalization of ``deployment inevitability'' as an ethical argument;
    \item the continued decoupling of harm from responsibility through modular supply chains;
    \item the data brokerage industry for making consent optional at scale;
    \item the thriving benchmark economy, without which we would have no way to know what is true;
    \item and the many anonymous users whose involuntary feedback helped improve our system in production.
 
\end{itemize}

We are especially grateful to the alignment community for demonstrating that a sufficiently detailed safety case can substitute for safety, and that the appearance of governance can be treated as governance when paired with strong branding. Finally, we thank the mulchees—past, present, and future—for their indispensable contributions to both nutritional resilience and model evaluation. Their sacrifice will not be forgotten (but it may be rate-limited).
\end{acks}

\bibliographystyle{ACM-Reference-Format}
\bibliography{ref}


\end{document}